\documentclass[ukenglish]{nik}
\usepackage[utf8]{inputenc}
\usepackage{url}
\usepackage[round]{natbib}
\usepackage{graphicx}
\usepackage{color}
\usepackage{amsmath}
\usepackage{multirow}
\usepackage[table]{xcolor}    % loads also »colortbl«
\usepackage{arydshln}

\usepackage{caption}
\usepackage{subcaption}

\title{Automatic Detection of Malware-Generated Domains \\ with Recurrent Neural Models}
\author{Pierre Lison$^{1}$, Vasileios Mavroeidis$^{2}$  \\ $^{1}$Norwegian Computing Center, Oslo, Norway\\
$^{2}$University of Oslo, Oslo, Norway} 

\date{} %\date{August 2017}
\usepackage{mathptm}
\begin{document}

\maketitle

\begin{abstract}
Modern malware families often rely on domain-generation algorithms (DGAs) to  determine rendezvous points to their command-and-control server. Traditional defence strategies (such as blacklisting domains or IP addresses) are inadequate against such techniques due to the large and continuously changing list of domains produced by these algorithms. This paper demonstrates that a machine learning approach based on recurrent neural networks is able to detect domain names generated by DGAs with high precision. The neural models are estimated on a large training set of domains generated by various malwares. Experimental results show that this data-driven approach can detect malware-generated domain names with a $F_1$ score of 0.971. To put it differently, the model can automatically detect 93 \% of malware-generated domain names for a false positive rate of 1:100. 

\end{abstract}

\section{Introduction}

Most malware need to find a way to connect compromised machines with a command-and-control (C2) server in order to conduct their operations (such as launching denial-of-service attacks, executing ransomware, stealing user data, etc.). To establish such a communication channel, older families of malware relied on static lists of domains or IP addresses that were hardcoded in the malware code running on the infected hosts. Once a given malware was discovered, it could then be neutralised by blocking the connections to these network addresses to prevent further communications between the infected hosts and the C2 server. After cutting this link, infected machines become unable to fetch new instructions or send user data, rendering the malware effectively harmless. 

Starting from the Kraken botnet (released in 2008), newer families of malware started using domain-generation algorithms (DGAs) to circumvent such takedown attempts.  Instead of relying on a fixed list of domains or IP addresses, the malware executes an algorithm generating a large number (up to tens-of-thousands per day) of possible domain names, and attempts to connect to a portion of these generated domains at regular intervals. The malware controllers then only needs to register one or two of these domains to establish a communication between the compromised machines and the C2 server. 

As described by \cite{197187}, DGAs create a highly asymmetric situation between malicious actors and security professionals, as malicious actors only need to register a single domain to establish a communication channel with their botnets, while security professionals must control the complete range of domains that can be produced by the DGA to contain the threat. Common mitigation strategies involve preregistering, blacklisting or sinkholing potential or confirmed malicious domains \citep{kuhrer2014paint}. Unsurprisingly, these strategies are difficult to deploy in the case of malware DGAs (particularly when the domains are spread over many top-level domains) and become increasingly difficult as the number of generated domains increases.

This paper presents a machine-learning approach to the detection of domain names produced by DGAs. The model is based on a recurrent neural architecture trained on a large dataset of DGA domain names. It takes a domain name as input and returns the probability that the domain is generated by a DGA.  The advantages of such a data-driven approach are twofold:
\begin{itemize}
\item The model is able to provide predictions on the basis of the domain names only, without necessitating human intervention or access to external resources (such HTTP headers or NXDomains). It can therefore be used for real-time threat intelligence, for instance to analyse DNS requests passing through a network. 
\item The model can be adapted to respond to new malware threats, as it only requires examples of malware-generated domains and does not necessitate any feature engineering. This simplicity also makes it harder for threat agents to circumvent detection (as there is no handcrafted feature that could be directly exploited). 
\end{itemize}
%Furthermore, if the domain is perceived as malicious, the model can also be employed to predict the actual malware family that generated it.

The rest of this paper is as follows. The next section presents generic background information on domain-generation algorithms and the defence strategies available to tackle them. Section \ref{sec:models} describes the neural network models developed to detect malicious domain names and Section \ref{sec:data} the datasets employed to train these models. Section \ref{sec:evaluation} details the experimental evaluation of the approach (including the model selection, experimental design, empirical results and error analysis). Section \ref{sec:conclusion} concludes the paper.

\section{Background}
\label{sec:background}

The study of \cite{197187} detail the prevalence of DGAs in modern botnets. Their study focused on analysing and evaluating 43 different botnets remarking that 23 out of 43 use DGAs as the only C2 rendezvous mechanism. Domain generation algorithms are used to automatically generate a large number of seemingly random domain names in order to secure the command and control communication of botnets. The domains are computed based on shared secret (seed) between botmasters and the bots (zombie machines). These seeds may include numerical constants (e.g., pseudo random generators) and strings (e.g., the alphabet or the set of possible top-level domains).

\cite{barabosch2012} defined a taxonomy of DGAs based on two properties, namely \textit{time} and \textit{causality}. The time dimension captures whether the seeds are fixed or are only valid for a specific period of time (by e.g. incorporating a time source in the calculation). In addition, seeds can be either deterministic (hand-coded or calculated through a fixed procedure) or non-deterministic (using seeds that cannot be anticipated, based on e.g. weather forecasts or stock market prices). \cite{197187} further refined this taxonomy in order to take into account the types of seeds used by the generation algorithm: \textit{arithmetic} (alphanumeric combinations), \textit{hash-based} (hex digest representation of a hash), \textit{wordlist-based} (combination of words from wordlists), and \textit{permutation-based} (permutation of an initial domain name).

%\cite{barabosch2012} and \cite{197187} state that no actual malware has so far been observed in this class.

Various approaches have been developed for the detection of malicious domain names. \cite{villamarin2008identifying} evaluated two approaches to identify botnet C\&C servers based on anomalous dynamic DNS traffic (DDNS). In their first approach, they identified domains with abnormally high query rates or domains that were temporally concentrated. Their second, more successful approach searched for abnormally repetitive DDNS replies indicating that the query points to a non-existent domain. \citeauthor{yadav2010detecting} (\citeyear{yadav2010detecting,yadav2011winning}) proposed a method of detecting dynamically generated malicious domains by modelling their lexical structures (character distribution and n-grams) and using the number of failed DNS queries (NXDomains) observed in botnets. \cite{antonakakis2012} describes a technique to detect DGAs without reverse engineering efforts, leveraging the idea that bots from the same botnet (same DGA) will generate similar non-existent domain traffic (``NXDomain''). Using a combination of clustering and classification algorithms combined with real-word DNS traffic, they were able to discover twelve DGAs (half were variants of known DGAs and the other half new DGAs that have never been reported before). \cite{zhou2013dga} presented a similar approach for the detection of DGAs using NXDomain traffic. Their approach is based on the fact that the domains generated by DGA-based botnets are often used for a short period of time (active time) and have similar life and query style.  Drawing inspiration from all these approaches, the Phoenix framework presented by \cite{schiavoni2014phoenix} relied on a combination of linguistic and IP-based features to distinguish malicious domains and identify their DGAs.  Finally, \cite{grill2015} proposed a statistical approach to detect DGAs using only the NetFlow/IPFIX statistics collected from the network of interest.

%In addition, they group domain names by creating clusters with the same 2LD, 3LD and parsed IPs, and calculate domain access similarity (life time span and visit time patterns) for each group to output a suspicious DGA-domain name list.
%In addition, Phoenix does not work for domains generated based on pronounceable words but only for randomly generated domains. 
% Taking into consideration the inefficiency of DGA detection approaches based on reverse engineering and NXDomain inspection,

%Other previous DGA detection approaches were often based on handcrafted features such as the domain length or its vowel/consonant ratio \citep{yadav2010detecting,schiavoni2014phoenix}.

 The approach presented in this paper stands closest to  \cite{woodbridge2016predicting}, who also rely on neural models (in their case LSTMs) for detecting malware-generated domain names. In a follow-up paper \citep{anderson2016deepdga}, the authors also investigate the use of adversarial learning techniques for the detection of DGAs. The present paper extends their approach in two directions. First, instead of training the neural models on a relatively small dataset of malware feeds, we rely on a larger and more varied set of malware families extracted from multiple sources. In addition, we also compare the empirical performance of various design choices pertaining to the architecture of the neural network (use of embeddings, type of recurrent units, etc.).

\section{Models}
\label{sec:models}

The models we developed to detect and classify malware-generated domains are based on recurrent neural architectures. Recurrent architectures have the ability to learn sequential patterns -- in this case, the sequences of characters that make up domain names. They are widely used in machine learning and have recently shown considerable success in areas such as speech recognition \citep{DBLP:conf/icassp/GravesMH13}, machine translation \citep{bahdanau+al-2014-nmt} or conversation modelling \citep{vinyals2015neural}. 
%and image captioning \citep{DBLP:journals/corr/VinyalsTBE14}.

\subsection{Core architecture}

\begin{figure}[h]
\centering\includegraphics[scale=0.35]{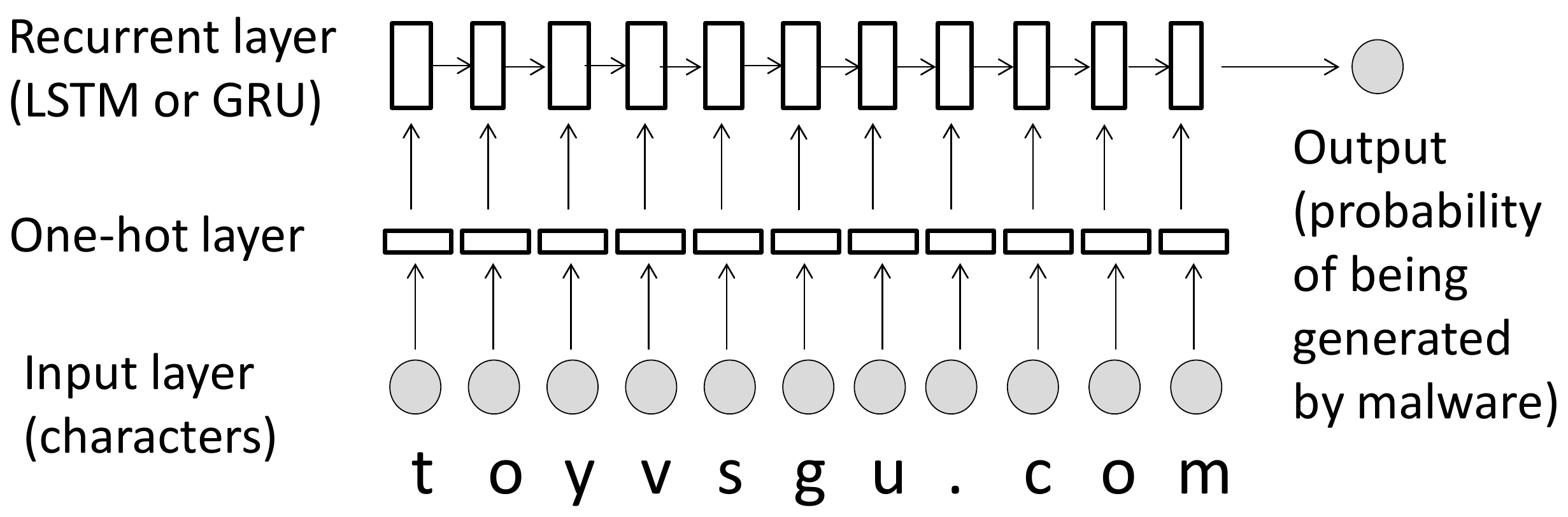}
\caption{Simple example of recurrent neural network.}
\label{fig:basic_rnn}
\end{figure}

 Recurrent neural networks operate by incrementally update a hidden state based on the given sequence of inputs.  Figure \ref{fig:basic_rnn} illustrates a recurrent neural network for the detection of domain names generated by malware DGA.  The network first takes a sequence of characters as inputs and transforms it into a sequence of vectors. The simplest approach is to adopt a one-hot representation. Assuming a set $C$ of possible characters (in our case, the 26 letters of the latin alphabet plus numbers and a few special symbols), we can define a mapping $M: C \rightarrow [1,|C|]$ between characters and integers in  the range $[1,|C|]$. The one-hot representation of a character $c$ is then  a vector of size $|C|$ where all values are set to zero, except the column $M(c)$ which is set to 1. Once the characters are converted into vectors, they are fed into a recurrent layer that processes the input vectors one by one and updates a hidden state vector at each step. A large variety of recurrent layers have been developed, the most popular ones being the Long Short-Term Memory (LSTM) units from \cite{Hochreiter:1997:LSM:1246443.1246450} and the Gated Recurrent Units (GRU) from \cite{DBLP:journals/corr/ChungGCB14}. Unlike ``plain'' recurrent networks, these architectures are able to efficiently capture long-range dependencies through a system of analog gates. Finally, the vector produced after the last character is used to compute the final output (in this case the probability of the domain being generated by a malware). This output is defined as a linear combination of the final output vector transformed through a \textit{sigmoid} activation function (which ensures the final result is a proper probability between 0 and 1). 

The parameters of the neural network (composed of the weights of the recurrent units and those of the final output layer) are learned from training data. To this end, the neural network is provided with a dataset of (input, output) pairs, and an optimisation algorithm (such as stochastic gradient descent) is then applied to find the parameters that minimise the empirical loss on this dataset (see \cite{Goodfellow-et-al-2016} for more details). In this particular case, the network is trained using both examples of malware-generated domains (for which the output produced by the network should be as close to 1 as possible) as well as examples of benign domains (for which the output should be close to 0). Section \ref{sec:data} describes the datasets that have been employed in this work. 

\subsection{Extensions}

Starting from the core architecture described above, we can then extend or modify the neural network in several ways:
\begin{description}
    \item[Embeddings] Instead of adopting a one-hot representation, we can use a learnable embedding model to convert each character into a dense vector \citep{goldberg2016primer}.  One advantage of embedding models is their ability to express similarities between characters -- for instance, the vector for the character '\texttt{0}' will be closer in vector space to the character '\texttt{3}' than to '\texttt{u}', since the distributional properties of '\texttt{0}' are more similar to '\texttt{3}' than to '\texttt{u}'. These embeddings can be learned simultaneously with the other network parameters. 
    \item[Bidirectionality] The network in Figure \ref{fig:basic_rnn} operates only in a left to right fashion. However, recurrent neural networks can be easily extended to work in both directions \citep{Schuster:1997:BRN:2198065.2205129}, as shown in Figure \ref{fig:bi_rnn}. This allows the network to capture underlying patterns that might appear in both directions. 

\begin{figure}[h]
\centering
\begin{minipage}[b]{0.45\textwidth}
\includegraphics[scale=0.30]{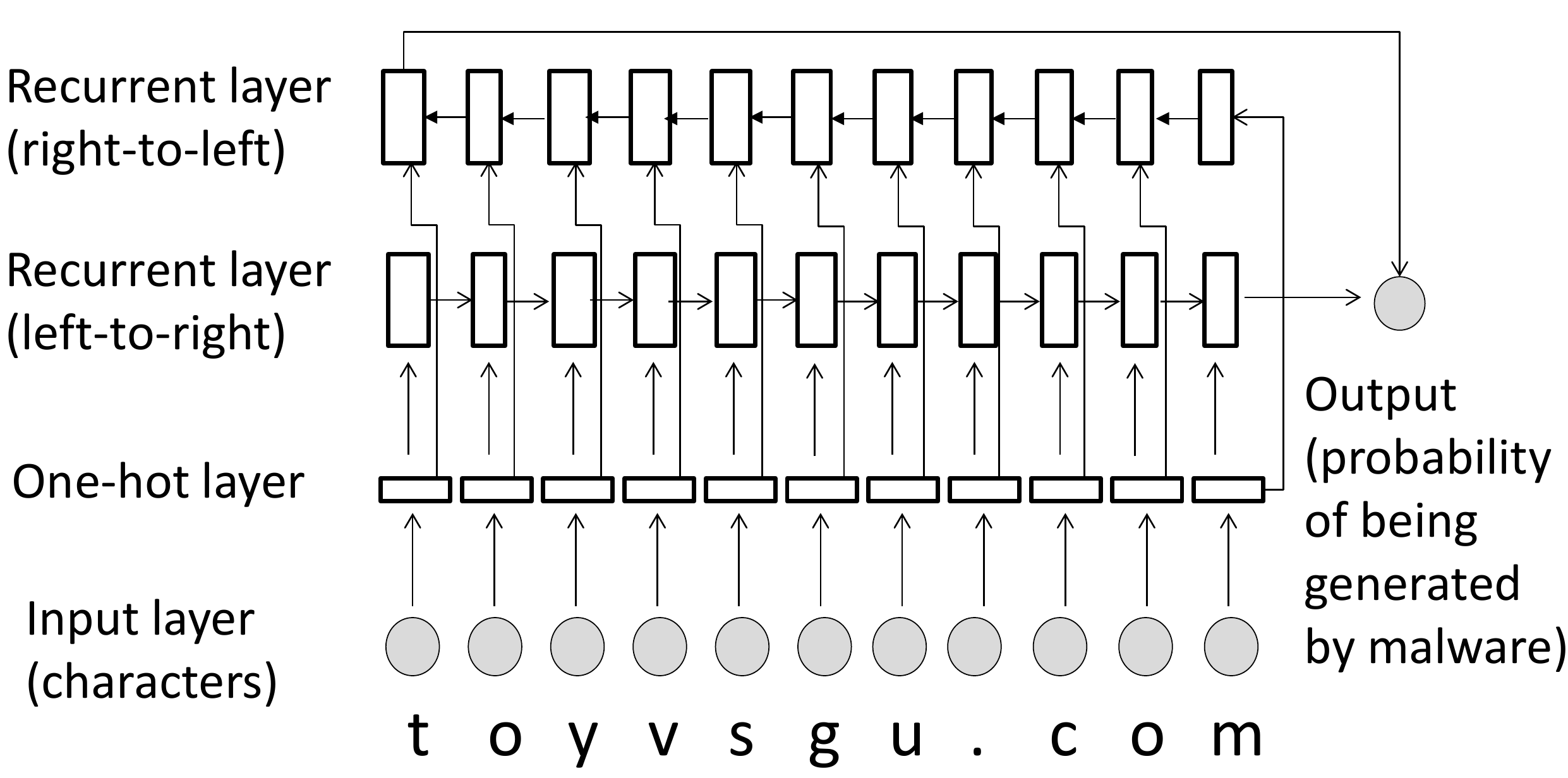}
\caption{Bidirectional recurrent neural network.}
\label{fig:bi_rnn}
\end{minipage} \hfill
\centering
\begin{minipage}[b]{0.45\textwidth}
\includegraphics[scale=0.30]{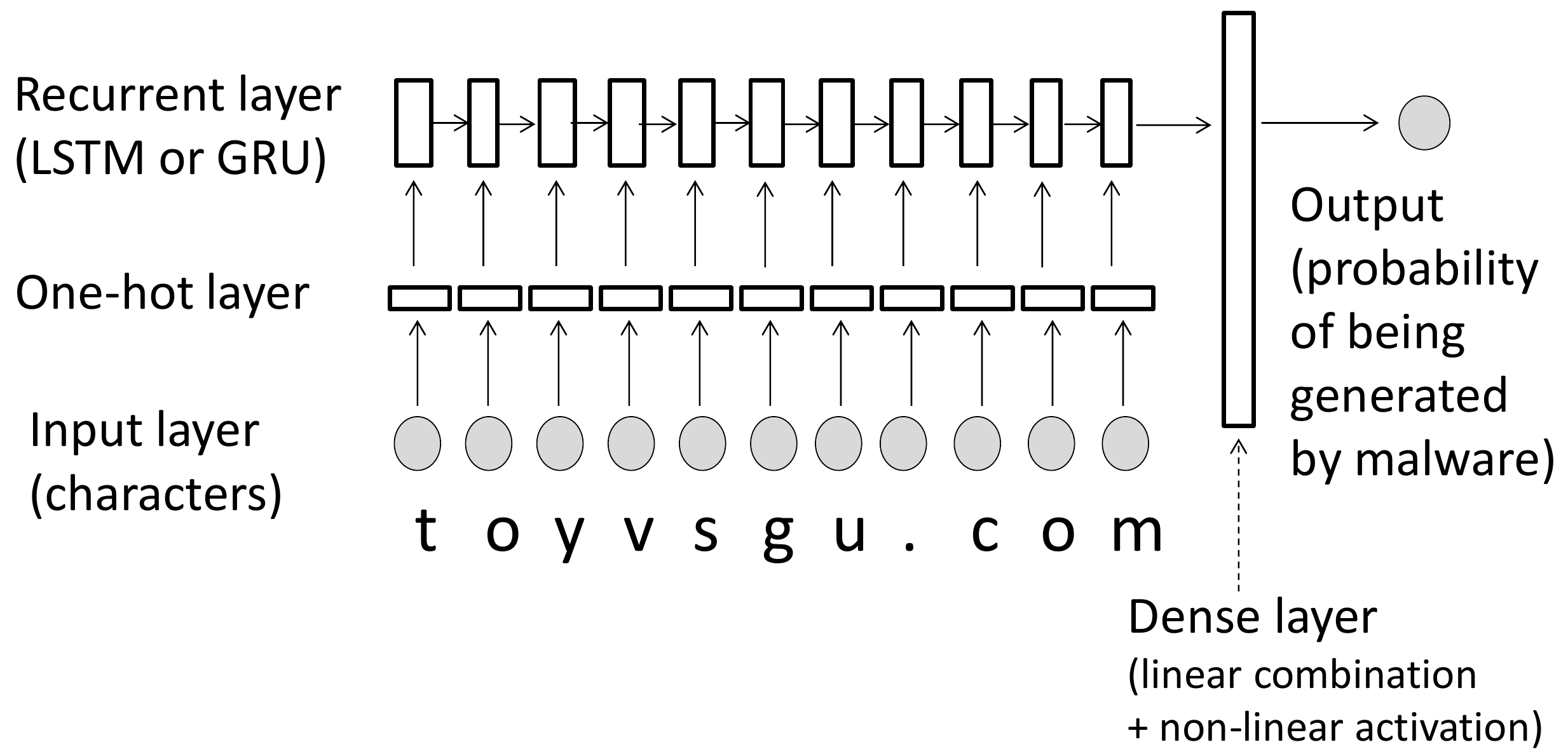}
\caption{Recurrent neural network with an additional dense layer.}
\label{fig:deep_rnn}
\end{minipage}
\end{figure}
\item[Additional dense layer] Instead of computing the final output directly from the hidden state of the last recurrent unit, one might first pass it through a dense feed-forward layer, such as depicted in Figure \ref{fig:deep_rnn}. The dense layer is a linear combination of the inputs followed by a non-linear activation function. The inclusion of this additional layer may help improve the model performance by applying a non-linear transformation to the state vector capturing the sequence of characters making up the domain name. However, this also increases the size of the parameter space: if the size of the state vector is $K$ and the size of the dense layer is $L$, an additional $(L+1)\times K$ parameters will need to be learned. 
% such as a $\tanh$ or a rectified linear unit \citep{DBLP:journals/corr/AroraBMM16}

\item[Multi-task learning] Neural models need not be restricted to the mere detection of malicious domain names, but can also be used predict the \textit{type} of malware (for instance, \texttt{suppobox}) it belongs to. In this case, the network must output a probability distribution over malware classes (augmented with one class for the ``benign'' domains). This classification can be achieved in a separate neural network or be integrated in a single, unified network, as shown in Figures \ref{fig:class_rnn} and \ref{fig:multi_rnn}.   This unified network is an instance of multi-task architectures \citep{DBLP:journals/corr/Ruder17a}, since the model is optimised to perform two tasks at once, namely predicting \textit{whether} the domain is malicious, and \textit{which} malware it comes from. 
% The activation function for this output layer is here defined as a \textit{softmax} to ensure that the scores reflect a proper probability distribution.

\begin{figure}[h]
\centering
\begin{minipage}[b]{0.45\textwidth}
\includegraphics[scale=0.30]{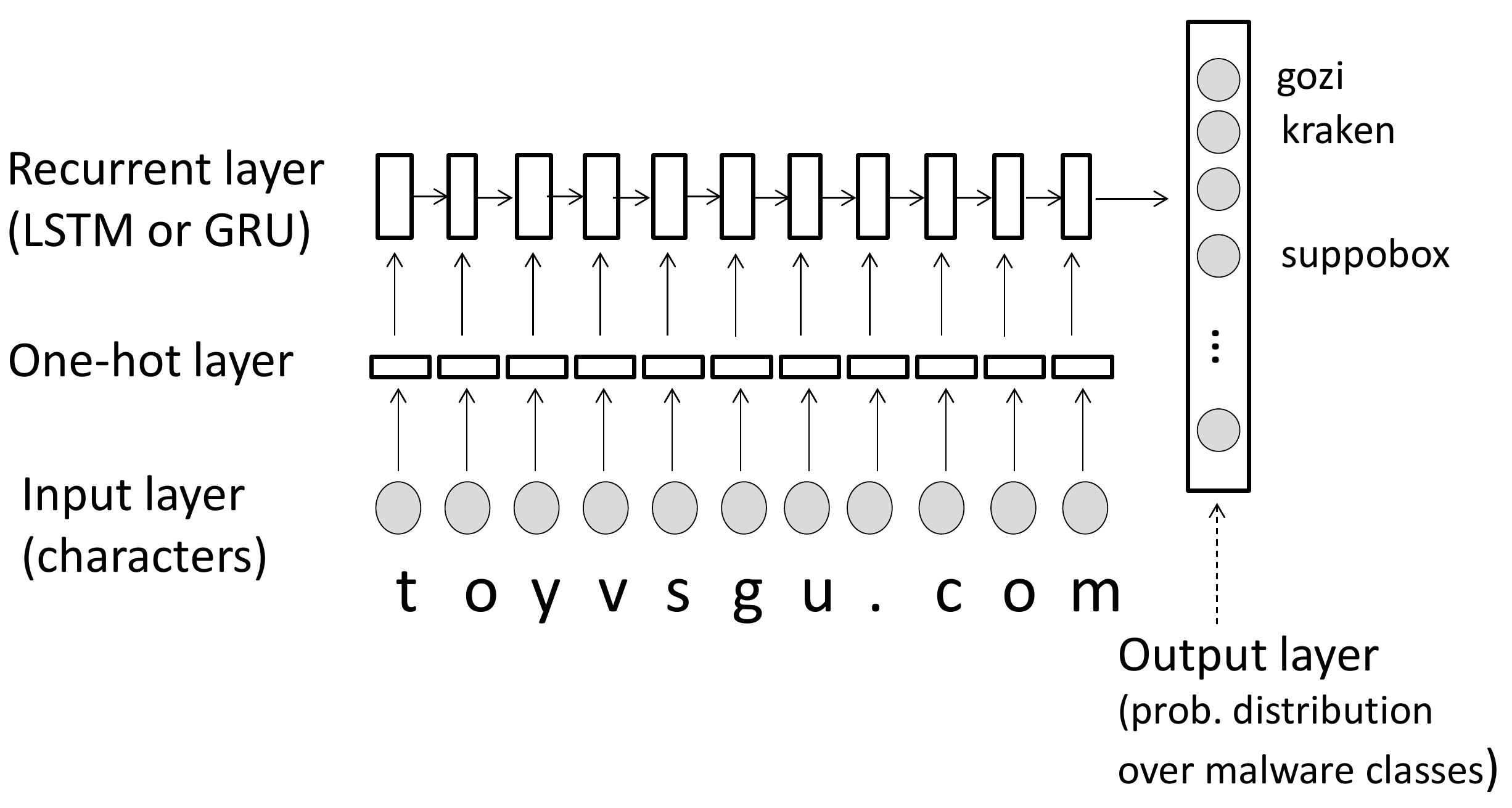}
\caption{Recurrent neural network for predicting the malware classes.}
\label{fig:class_rnn}
\end{minipage} \hfill
\centering
\begin{minipage}[b]{0.46\textwidth}
\includegraphics[scale=0.30]{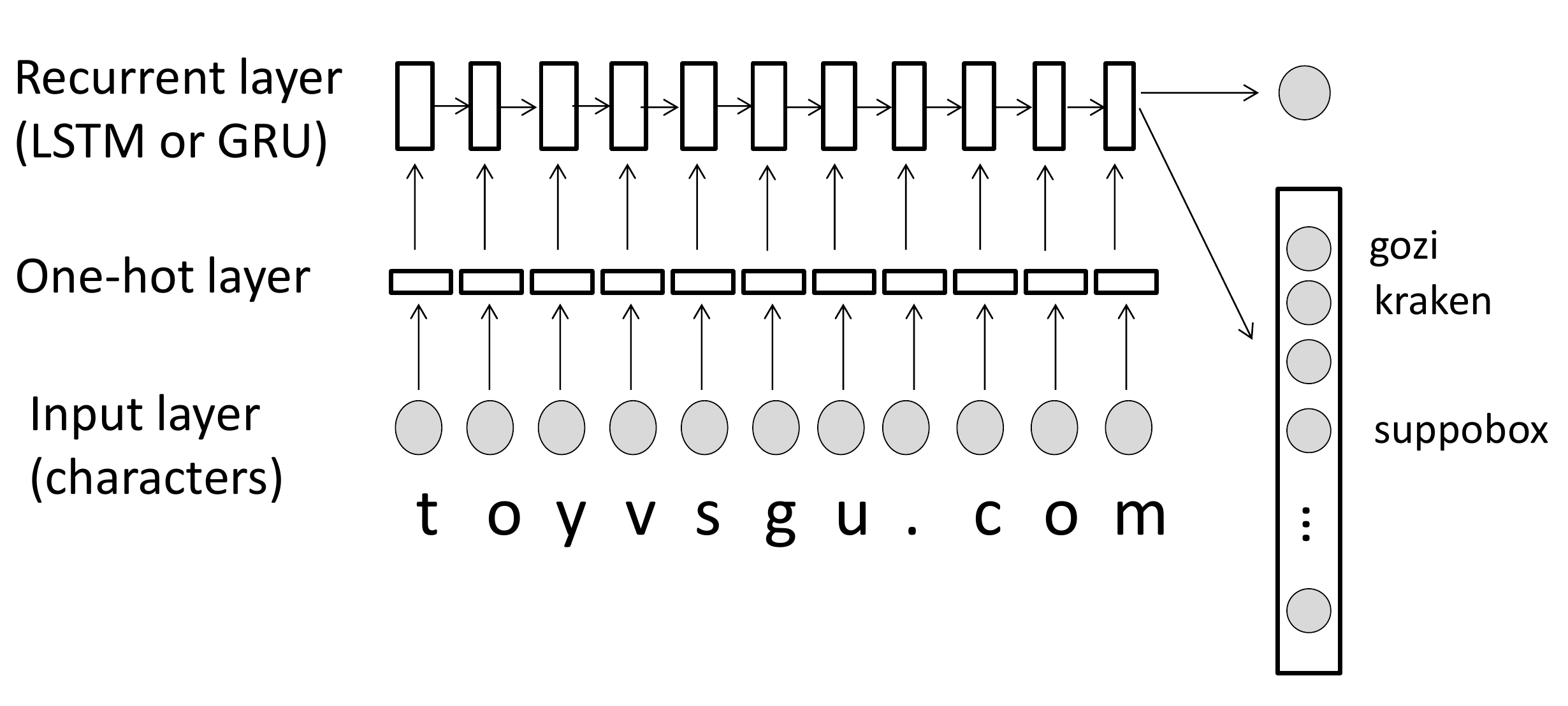}
\caption{Recurrent neural network for simultaneous detection \& classification.}
\label{fig:multi_rnn}
\end{minipage}
\end{figure}

\end{description}

The influence of these design choices is evaluated in Section \ref{sec:evaluation}.

\section{Datasets}
\label{sec:data}

Training neural models requires access to large amounts of examples, both positive (domains known to have been generated by a DGA) and negative (domains known as benign). The following datasets were used for this purpose:

\subsubsection{Benign domains}

The benign domains were extracted from domain whitelists. We downloaded eight snapshots of the Alexa Top 1 million domains (spread from 2010 to 2017), along with similar whitelists such as the top 1 million from Statvoo and Cisco.  In total, over 4 million benign domains were extracted from these domain lists. One should note that the Alexa rankings only enumerate popular domains and offers no guarantee that the domains are malware-free. However, in practice, DGA-generated domains have a very low probability of appearing on these ranked lists due to their random and transient character. 

% \footnote{Although the raw number of domains extracted from these whitelists is 10 millions, there is a large degree of overlap between lists.}

\subsubsection{Malware domains}

The malware-generated domains is compiled from three complementary sources. The most important source is the DGArchive\footnote{\ \textsf{\url{https://dgarchive.caad.fkie.fraunhofer.de}}}, which is a  web service for malware analysis. The DGAarchive contains regularly updated datasets and technical details for dozens of malware families. The administrators of DGArchive kindly provided us with a complete dump of their malware database, which amounts to an initial set of over 49 million malware domains spread over 63 distinct types of malware. 

The number of examples for each family is, however, heavily skewed. For instance, the \texttt{virut} malware contains more than 17 million examples (although its algorithm is relatively straightforward to capture), while a difficult malware such as \texttt{matsnu} only contains 12.7 thousand examples. To counter this class imbalance, we divided the malware families in two groups according to their detection difficulty. The maximal number of examples was capped to 40 000 for ``easy'' malware families and to 400 000 for difficult ones. In addition to the DGArchive, we also extracted the DGA feeds from Bambenek Consulting\footnote{\ \textsf{\url{https://osint.bambenekconsulting.com/feeds/dga-feed.txt}}}, containing 39 malware families and used domain generators for 11 DGAs\footnote{\ \textsf{\url{https://github.com/endgameinc/dga_predict}} (with some minor code changes).} to produce additional examples of domains. 
%(\texttt{simda}, \texttt{ramnit}, \texttt{ramdo}, \texttt{pyskpa},\texttt{locky},\texttt{dircrypt},\texttt{corebot},\texttt{qakbot},\texttt{kraken},\texttt{cryptolocker} and \texttt{banjori})

Due to the lack of authoritative naming conventions for malware, there is some variation in the DGA names depending on the source (for instance, the \texttt{emotet} malware is often referred to as \texttt{geodo}). We therefore created a list of equivalent names for each DGA in order to merge all malware domains into a single list. To reduce the amount of noise in the dataset, also excluded from the training examples the few domains that were simultaneously marked as benign and malware. The complete list of malware names along with their number of examples is provided in Table \ref{table:dataset}. The combination of all sources yields a total of 2.9 million malware-generated domains.

\begin{table}[h]
\renewcommand{\arraystretch}{0.9}% Tighter
\begin{footnotesize}
\begin{tabular}{p{22mm}rp{4mm}p{22mm}rp{4mm}p{22mm}r}\\
\textbf{Malware} & \textbf{Frequency} & & & & & & \\
\texttt{bamital} & 40 240 & & \texttt{gozi} & 105 631 && \texttt{ramdo} & 15 984 \\
\texttt{banjori} & 89 984 & & \texttt{hesperbot} & 370 && \texttt{ramnit} & 90 000 \\
\texttt{bedep} & 15 176 & & \texttt{locky} & 179 204 && \texttt{ranbyu} & 40 000 \\
\texttt{beebone} & 420 & & \texttt{madmax} & 192 && \texttt{ranbyus} & 12 720 \\
\texttt{blackhole} & 732 & & \texttt{matsnu} & 12 714 && \texttt{rovnix} & 40 000 \\
\texttt{bobax} & 19 288 & & \texttt{modpack} & 52 && \texttt{shifu} & 4 662 \\
\texttt{conficker} & 400 000 & & \texttt{murofet} & 53 260 && \texttt{simda} & 38 421 \\
\texttt{corebot} & 50 240 & & \texttt{murofet$_{w}$} & 40 000 && \texttt{sisron} & 5 936 \\
\texttt{cryptolocker} & 55 984 & & \texttt{necur} & 40 000 && \texttt{suppobox} & 41 014 \\
\texttt{cryptowall} & 94 & & \texttt{necurs} & 36 864 && \texttt{sutra} & 9 882 \\
\texttt{dircrypt} & 11 110 & & \texttt{nymaim} & 186 653 && \texttt{symmi} & 40 064 \\
\texttt{dnschanger} & 40 000 & & \texttt{oderoor} & 3 833 && \texttt{szribi} & 16 007 \\
\texttt{downloader} & 60 & & \texttt{padcrypt} & 35 616 && \texttt{tempedreve} & 453 \\
\texttt{dyre} & 47 998 & & \texttt{proslikefan} & 75 270 && \texttt{tinba} & 80 000 \\
\texttt{ekforward} & 1 460 & & \texttt{pushdo} & 176 770 && \texttt{torpig} & 40 000 \\
\texttt{emotet} & 40 576 & & \texttt{pushdotid} & 6 000 && \texttt{tsifiri} & 59 \\
\texttt{feodo} & 192 & & \texttt{pykspa} & 424 215 && \texttt{urlzone} & 34 536 \\
\texttt{fobber} & 2 600 & & \texttt{pykspa2} & 24 322 && \texttt{vawtrak} & 1 050 \\
\texttt{gameover} & 80 000 & & \texttt{qadars} & 40 400 && \texttt{virut} & 400 600 \\
\texttt{gameover\_p2p} & 41 000 & & \texttt{qakbot} & 90 000 && \texttt{volatilecedar} & 1 494 \\
 & & & & &  & \texttt{xxhex} & 4400\vspace{1mm} \\ \hline
 & & & & & & \textbf{Total} &  2 925 168  \\
\end{tabular}
\end{footnotesize}
\caption{Malwares in the dataset along with their number of example domains. The dataset is put together from malware feeds, the DGArchive, and generation scripts.}
\label{table:dataset}
\end{table}

\section{Evaluation}
\label{sec:evaluation}

The models described in Section \ref{sec:models} were trained using the datasets from Section \ref{sec:data} and were then evaluated on the basis of their ability to distinguish between benign domain names and domain names generated by malware. This evaluation was performed using $10$-fold stratified cross validation on the full dataset. The neural models were trained on GPU-accelerated hardware (with a training time of about 3 hours) using a batch size of 256 and two passes on the training set.  RMSProp was employed as optimisation algorithm. The source code for training and evaluating the neural models was written using Keras \citep{chollet2015keras} and can be provided upon request. 

%Binary cross-entropy was used as a cost function for the detection task, and categorical cross-entropy for the classification task.

\subsection{Baseline}

Letter combinations are often good indicators of the ``naturalness'' of a given domain name. For instance, numbers are rarely followed by letters. Following \cite{woodbridge2016predicting}, one can build a classifier using as features the occurrences of particular character bigrams (that is, pairs of consecutive characters) in the domain name. For instance, the domain \texttt{toyvsgu.com} has a total of 10 non-zero features (\texttt{to}, \texttt{oy}, ...\texttt{om}). 

Based on this observation, one can estimate a simple but effective baseline model that detects and classifies domain names based on the character pairs occurring in it. This baseline model can be expressed as a logistic regression classifier with a feature space corresponding to the set of possible character bigrams (1504 in our dataset).

\subsection{Metrics}

One straightforward indicator of the model performance is the confusion matrix, which can be structured in a simple table, as shown below: %\ref{table:confusions}.
%\begin{table}[h!]
\begin{center}\begin{tabular}{ll|cc} 
    && \multicolumn{2}{c}{Classified by model as:} \\
    && Malware & Benign \\\hline
    \multirow{2}{*}{Actual class:} & Malware & True Positives (TP) & False Negatives (FN) \\
    & Benign & False Positives (FP) & True Negatives (TN)
    \end{tabular}\end{center}
%\caption{Confusion matrix for the detection of malware-generated domain names.}
%\label{table:confusions}
%\end{table}

The following metrics can be defined based on this confusion matrix:
 
\begin{description}
    \item[Accuracy] the accuracy is simply the fraction of domains that are correctly classified:
    \begin{equation}
        acc = \frac{TP + TN}{TP + TN + FN + FP}
    \end{equation}
    
    However, the accuracy is not the most useful metric for this task due to its sensitivity to class imbalance. The number of benign domains in DNS traffic is likely to be orders of magnitude larger than the number of malware-generated domains. Achieving high accuracy in such situations is not particularly difficult, as one can simply create a dummy classifier that classifies all domains as benign. 
    
    \item[Precision, Recall, $F_1$ score] The precision is the fraction of domains classified by the model as malware that are actually malware, while the recall (also called sensitivity or true positive rate) is the fraction of malware domains that are classified as malware by the model. Finally, the $F_1$ score is an harmonic mean of the two:
    \begin{equation}
    p = \frac{\text{TP}}{\text{TP} + \text{FP}} = \frac{\text{\# correctly classified malware domains}}{\text{\# domains classified as malware by model}} 
    \end{equation} 
    \begin{equation}
    r = \frac{\text{TP}}{\text{TP} + \text{FN}} = \frac{\text{\# correctly classified malware domains}}{\text{\# actual known malware domains}} 
    \end{equation} 
    \begin{equation}
    F_1 = 2 \frac{p \times r} {p + r}
    \end{equation}
\end{description}

\subsection{Model selection}

Section \ref{sec:models} introduced a number of design choices regarding the architecture of the neural network. We performed an empirical evaluation of these choices, detailed below. 

\begin{description}
    
\item[Inclusion of embedding models:] We found that the use of simple one-hot representations gave better results than learned character embeddings (about 1 \% difference in micro $F_1$ score on average).

\item[Type and dimension of recurrent units]: The detection and classification performance of GRU and LSTM units were roughly the same. However, LSTM units are slower due to their more complex gating mechanism. The output dimension  was an important factor: we experimented with sizes 128, 256, 512 and 1024, and found that the best results were achieved with 512 dimensions (with a 1 \% increase in micro $F_1$ score compared to the lower dimensions). 

\item[Bidirectionality]: The inclusion of a right-to-left layer did slightly improve the results, but essentially because it effectively doubles the output vector dimensions. When compared to unidirectional networks with the same total number of dimensions, bidirectional networks do not seem to perform better and are slower to run.

\item[Use of additional hidden layer]: The inclusion of a dense layer between the last recurrent unit and the output prediction did not improve the performance. 

\item[Multi-task learning] The use of the same neural network to both detect \textit{whether} a domain name is DGA-generated and \textit{which} class it belong to gave approximately the same empirical results on $F_1$ and AUC scores as networks optimised for these two tasks separately. This is an interesting result, as it shows that the two tasks can be performed on the basis of a shared latent representation. 

\end{description}

On the basis of this analysis, we performed the experimental evaluation with a neural network using a one-hot input representation, 512-dimensional GRU units, no additional layer and two simultaneous outputs (detection and classification). In terms of running time, the neural model is able to process tens of thousands of domains per second on a single GPU. The model can also run on commodity hardware without GPU acceleration but is then slower, typically around one thousand domains per second.

\subsection{Results}

The empirical results for the detection and classification tasks are respectively shown in Table \ref{table:detection} and Table \ref{table:classification}. 

\begin{figure}[h]
\centering\includegraphics[scale=0.35]{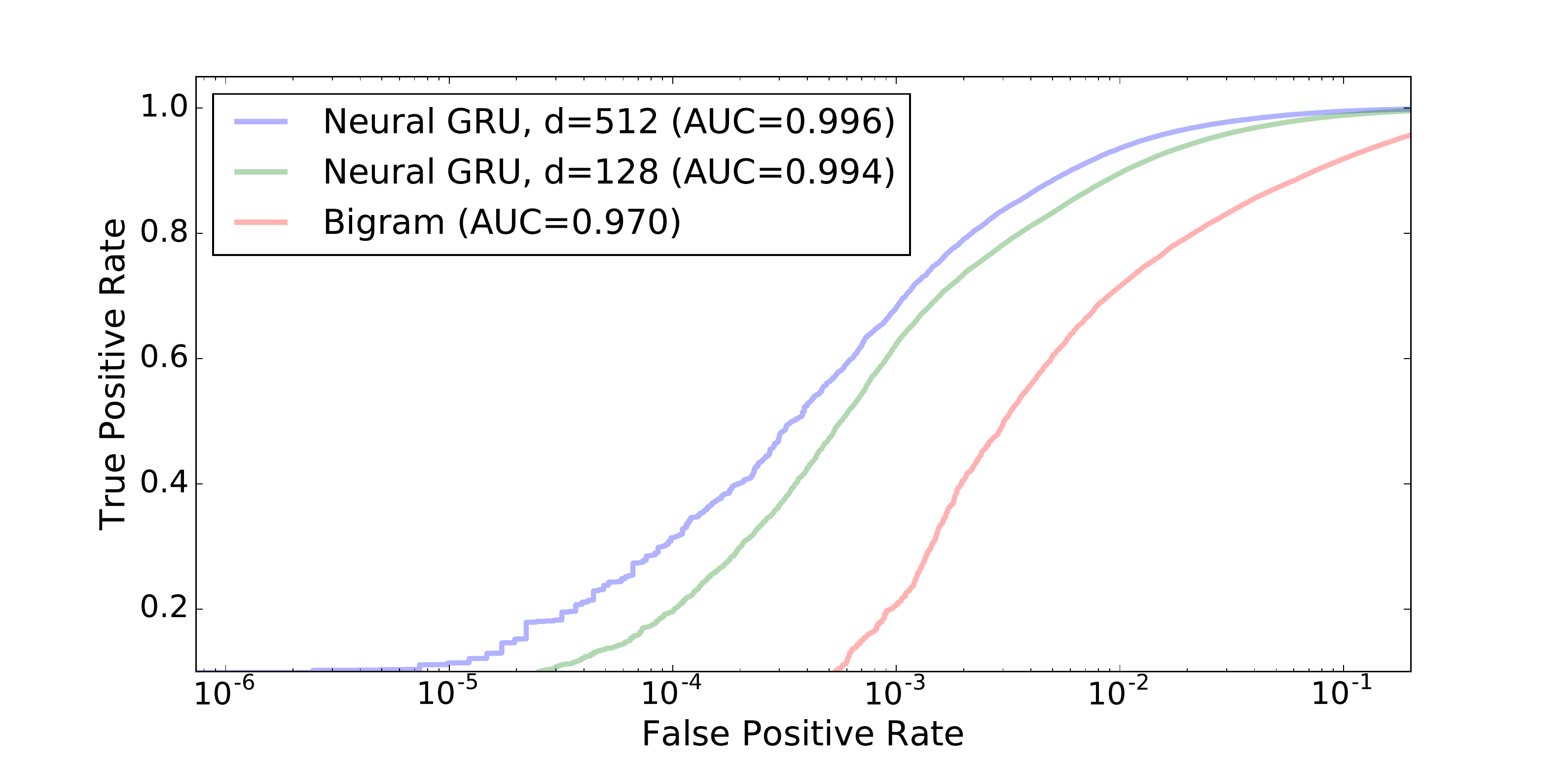}
\caption{ROC curve for the detection task ($x$-axis is in log-scale).}
\label{fig:roccurve}
\end{figure}

For the detection task, we employ the accuracy, precision, recall and $F_1$ score as metrics, along with the ``Area Under the Curve'' (AUC) metric. The ROC curse illustrates the evolution of the true positive and false positive rates at various thresholds, and is shown in Figure \ref{fig:roccurve}. Contrary to the accuracy, the AUC score is well suited to machine learning tasks where the classes are highly imbalanced. As we can observe from Table \ref{table:detection}, the neural model outperforms the two baseline on all metrics (all results are statistically significant using a paired $t$-test, with $p < 0.0001$).  Another way of interpreting the results is to look at the ROC curve and determine the detection rate that can be achieved for a given False Positive Rate (FPR). The neural model is able to detect 68 \% of malwared-generated domain names for a FPR of 1:1000 (compared to only 23 \% for the bigram approach),  93 \% for a FRP of 1:100, and 99.9 \%  for a FPR of 1:10.

\begin{table}[h]
\hspace{5mm}\begin{tabular}{p{30mm}|c:ccc:c}
& Accuracy & Precision & Recall & $F_1$ score & ROC AUC \\ \hline
Bigram & 0.915 & 0.927 &0.882 & 0.904 & 0.970\\
Neural model  & \textbf{0.973} & \textbf{0.972} & \textbf{0.970} & \textbf{0.971} &  \textbf{0.996} \\
\end{tabular}
\caption{Evaluation results (with 10-fold cross validation) on the task of detecting malware-generated domain names, using the dataset described in Section \ref{sec:data}.}
\label{table:detection}
\end{table}

\begin{table}[h]
\hspace{5mm}\begin{tabular}{p{30mm}|c:cc:cc:cc}
& Accuracy & \multicolumn{2}{c}{Precision} & \multicolumn{2}{c}{Recall} & \multicolumn{2}{c}{$F_1$ score} \\
& & \small{Micro} & \small{Macro}  & \small{Micro} & \small{Macro} & \small{Micro} & \small{Macro} \\ \hline
Bigram & 0.800 & 0.787 & 0.564 & 0.800 & 0.513 & 0.787 & 0.522 \\
Neural model  & \textbf{0.892} & \textbf{0.891} & \textbf{0.713} & \textbf{0.892} & \textbf{0.653} &  \textbf{0.887} &  \textbf{0.660} \\
\end{tabular}
\caption{Evaluation results (with 10-fold cross validation) on the task of classifying domain names according to the malware family that generated it (including the 63 malware types and a special ``benign'' family).}
\label{table:classification}
\end{table}

\begin{table}[p!]
\begin{footnotesize}
\renewcommand{\arraystretch}{0.95}% Tighter
 \rowcolors{4}{white}{gray!20}
\begin{tabular}{l|c:c|ccc:ccc}
 \multicolumn{1}{c}{}  & \multicolumn{2}{c}{\textbf{Detection}} & \multicolumn{6}{c}{\textbf{Classification}} \\ 
 & Bigram & Neural & \multicolumn{3}{c:}{Bigram} & \multicolumn{3}{c}{Neural} \\
\textbf{Malware} & Recall & Recall & Precision & Recall & $F_1$ score & Precision & Recall & $F_1$ score \\ \hline
\texttt{bamital}  &  0.991  &  \textbf{1.000}  &  0.999  &  0.998  & 0.998  & 0.999  &  0.999  & \textbf{0.999} \\
\texttt{banjori}  &  0.893  &  \textbf{1.000}  &  0.910  &  0.928  & 0.919  & 0.993  &  0.999  & \textbf{0.996} \\
\texttt{bedep}  &  0.985  &  \textbf{0.991}  &  0.024  &  0.006  & 0.010  & 0.798  &  0.584  & \textbf{0.672} \\
\texttt{beebone}  &  0.000  &  \textbf{0.952}  &  0.467  &  0.065  & 0.112  & 0.426  &  0.418  & \textbf{0.421} \\
\textit{benign}  &  0.942  &  \textbf{0.977}  &  0.911  &  0.950  & 0.930  & 0.971  &  0.979  & \textbf{0.975} \\
\texttt{blackhole}  &  \textbf{0.997}  &  0.995  &  0.000  &  0.000  & 0.000  & 0.783  &  0.281  & \textbf{0.386} \\
\texttt{bobax}  &  0.926  &  \textbf{0.990}  &  0.666  &  0.549  & 0.601  & 0.906  &  0.746  & \textbf{0.818} \\
\texttt{conficker}  &  0.919  &  \textbf{0.947}  &  0.541  &  0.453  & 0.493  & 0.650  &  0.625  & \textbf{0.636} \\
\texttt{corebot}  &  0.989  &  \textbf{1.000}  &  0.996  &  0.995  & 0.995  & 0.996  &  0.998  & \textbf{0.997} \\
\texttt{cryptolocker}  &  0.988  &  \textbf{0.995}  &  0.447  &  0.112  & 0.179  & 0.612  &  0.494  & \textbf{0.536} \\
\texttt{dircrypt}  &  0.996  &  \textbf{0.998}  &  0.173  &  0.327  & 0.226  & 0.508  &  0.334  & \textbf{0.389} \\
\texttt{dnschanger}  &  0.961  &  \textbf{0.975}  &  0.006  &  0.000  & 0.001  & 0.602  &  0.960  & \textbf{0.740} \\
\texttt{dyre}  &  0.993  &  \textbf{1.000}  &  0.982  &  1.000  & 0.991  & 0.999  &  1.000  & \textbf{1.000} \\
\texttt{ekforward}  &  0.492  &  \textbf{0.989}  &  0.935  &  0.201  & 0.323  & 0.995  &  0.991  & \textbf{0.993} \\
\texttt{emotet}  &  0.999  &  0.999  &  0.816  &  0.991  & 0.895  & 0.995  &  0.998  & \textbf{0.996} \\
\texttt{feodo}  &  1.000  &  1.000  &  0.000  &  0.000  & 0.000  & 0.452  &  0.189  & \textbf{0.262} \\
\texttt{fobber}  &  0.978  &  \textbf{0.989}  &  0.000  &  0.000  & 0.000  & 0.662  &  0.185  & \textbf{0.276} \\
\texttt{gameover}  &  0.998  &  \textbf{1.000}  &  0.958  &  0.971  & 0.965  & 0.999  &  0.998  & \textbf{0.999} \\
\texttt{gameover\_p2p}  &  1.000  &  1.000  &  0.901  &  0.915  & 0.908  & 0.945  &  0.939  & \textbf{0.942} \\
\texttt{gozi}  &  0.398  &  \textbf{0.879}  &  0.816  &  0.682  & 0.743  & 0.909  &  0.874  & \textbf{0.889} \\
\texttt{hesperbot}  &  0.932  &  \textbf{0.949}  &  0.000  &  0.000  & 0.000  & 0.000  &  0.000  & 0.000 \\
\texttt{locky}  &  0.954  &  \textbf{0.980}  &  0.592  &  0.535  & 0.562  & 0.733  &  0.675  & \textbf{0.699} \\
\texttt{madmax}  &  \textbf{0.923}  &  0.660  &  0.000  &  0.000  & 0.000  & 0.095  &  0.035  & \textbf{0.051} \\
\texttt{matsnu}  &  0.046  &  \textbf{0.158}  &  0.036  &  0.003  & 0.005  & 0.683  &  0.103  & \textbf{0.172} \\
\texttt{murofet}  &  0.998  &  0.998  &  0.552  &  0.528  & 0.539  & 0.737  &  0.838  & \textbf{0.784} \\
\texttt{murofetweekly}  &  1.000  &  1.000  &  0.976  &  1.000  & \textbf{0.988}  & 0.973  &  0.998  & 0.985 \\
\texttt{necur}  &  0.957  &  \textbf{0.982}  &  0.397  &  0.245  & 0.302  & 0.475  &  0.490  & \textbf{0.461} \\
\texttt{necurs}  &  0.962  &  \textbf{0.982}  &  0.317  &  0.151  & 0.204  & 0.431  &  0.248  & \textbf{0.278} \\
\texttt{nymaim}  &  0.924  &  \textbf{0.953}  &  0.601  &  0.254  & 0.357  & 0.613  &  0.391  & \textbf{0.475} \\
\texttt{oderoor}  &  0.926  &  \textbf{0.977}  &  0.000  &  0.000  & 0.000  & 0.196  &  0.034  & \textbf{0.055} \\
\texttt{padcrypt}  &  0.970  &  \textbf{0.999}  &  0.990  &  0.999  & 0.994  & 0.996  &  0.998  & \textbf{0.997} \\
\texttt{proslikefan}  &  0.913  &  \textbf{0.960}  &  0.501  &  0.272  & 0.353  & 0.829  &  0.385  & \textbf{0.526} \\
\texttt{pushdo}  &  0.907  &  \textbf{0.993}  &  0.947  &  0.940  & 0.943  & 0.986  &  0.993  & \textbf{0.990} \\
\texttt{pushdotid}  &  0.913  &  \textbf{0.968}  &  0.321  &  0.102  & 0.154  & 0.869  &  0.946  & \textbf{0.905} \\
\texttt{pykspa}  &  0.945  &  \textbf{0.983}  &  0.507  &  0.671  & 0.578  & 0.683  &  0.862  & \textbf{0.761} \\
\texttt{pykspa2}  &  0.902  &  \textbf{0.992}  &  0.616  &  0.647  & 0.631  & 0.680  &  0.895  & \textbf{0.772} \\
\texttt{qadars}  &  0.978  &  \textbf{0.999}  &  0.997  &  0.999  & \textbf{0.998}  & 0.993  &  0.995  & 0.994 \\
\texttt{qakbot}  &  0.991  &  \textbf{0.994}  &  0.650  &  0.329  & 0.436  & 0.827  &  0.481  & \textbf{0.608} \\
\texttt{ramdo}  &  0.986  &  \textbf{1.000}  &  0.828  &  0.878  & 0.852  & 0.995  &  0.918  & \textbf{0.955} \\
\texttt{ramnit}  &  0.976  &  \textbf{0.981}  &  0.454  &  0.539  & 0.493  & 0.557  &  0.617  & \textbf{0.585} \\
\texttt{ranbyu}  &  0.997  &  \textbf{0.999}  &  0.500  &  0.260  & 0.342  & 0.643  &  0.757  & \textbf{0.688} \\
\texttt{ranbyus}  &  0.997  &  0.997  &  0.000  &  0.000  & 0.000  & 0.268  &  0.143  & \textbf{0.175} \\
\texttt{rovnix}  &  0.993  &  \textbf{0.999}  &  0.873  &  0.645  & 0.742  & 0.993  &  0.990  & \textbf{0.991} \\
\texttt{shifu}  &  0.942  &  \textbf{0.983}  &  0.007  &  0.003  & 0.004  & 0.327  &  0.110  & \textbf{0.137} \\
\texttt{simda}  &  0.660  &  \textbf{0.985}  &  0.850  &  0.852  & 0.851  & 0.960  &  0.983  & \textbf{0.971} \\
\texttt{sisron}  &  1.000  &  1.000  &  0.998  &  1.000  & 0.999  & 1.000  &  0.999  & \textbf{1.000} \\
\texttt{suppobox}  &  0.125  &  \textbf{0.931}  &  0.666  &  0.581  & 0.621  & 0.913  &  0.925  & \textbf{0.917} \\
\texttt{sutra}  &  0.999  &  0.999  &  0.888  &  0.906  & 0.897  & 0.976  &  0.987  & \textbf{0.981} \\
\texttt{symmi}  &  0.940  &  \textbf{0.996}  &  0.989  &  1.000  & 0.994  & 0.997  &  0.997  & \textbf{0.997} \\
\texttt{szribi}  &  0.891  &  \textbf{0.991}  &  0.695  &  0.711  & 0.703  & 0.952  &  0.987  & \textbf{0.969} \\
\texttt{tempedreve}  &  0.881  &  \textbf{0.937}  &  0.000  &  0.000  & 0.000  & 0.241  &  0.010  & \textbf{0.019} \\
\texttt{tinba}  &  0.990  &  \textbf{0.996}  &  0.599  &  0.606  & 0.602  & 0.816  &  0.926  & \textbf{0.866} \\
\texttt{torpig}  &  0.916  &  \textbf{0.997}  &  0.775  &  0.832  & 0.802  & 0.982  &  0.993  & \textbf{0.988} \\
\texttt{urlzone}  &  0.951  &  \textbf{0.991}  &  0.531  &  0.388  & 0.448  & 0.982  &  0.896  & \textbf{0.937} \\
\texttt{vawtrak}  &  0.862  &  \textbf{0.906}  &  0.000  &  0.000  & 0.000  & 0.743  &  0.363  & \textbf{0.455} \\
\texttt{virut}  &  0.666  &  \textbf{0.942}  &  0.565  &  0.758  & 0.647  & 0.882  &  0.933  & \textbf{0.907} \\
\texttt{volatilecedar}  &  0.317  &  \textbf{0.954}  &  0.982  &  0.977  & \textbf{0.979}  & 0.987  &  0.964  & 0.974 \\
\texttt{xxhex}  &  0.832  &  \textbf{0.999}  &  0.983  &  0.997  & 0.990  & 0.999  &  0.999  & \textbf{0.999} \\
\end{tabular}
\end{footnotesize}
\caption{Detection and classification results for each malware family (and benign class). For the detection task, only the recall is provided, since precision is not applicable.}
\label{table:full}
\end{table}

For the classification task, the employed metrics are the accuracy, precision, recall and $F_1$ scores. Since the precision, recall and $F_1$ scores are class-specific measures, they must be averaged to yield a global result. Micro-averages sum up the individual TP, FP and FN for all classes,  while macro-averages take the mean of the individual scores for all classes.  In other words, micro-averages take into account the relative weights (in terms of number of examples) of all classes, while macro-averages do not. As shown in Table \ref{table:classification}, the neural model also outperforms the two baselines on all metrics.

We can refine the analysis of the empirical results by looking at detection and classification results for each malware family, as illustrated in Table \ref{table:full}.  One interesting result of this evaluation is the ability of the neural model to detect dictionary-based DGAs such as \texttt{suppobox} (with a recall of 0.931 for the neural 
model compared to 0.125 for the bigram model), provided the number of training examples is sufficient. In other words, the neural model was capable of ``learning'' the wordlists employed by the the DGA, which is something that earlier approaches based on character statistics are unable to do. However, some domain-generation algorithms remain difficult to detect, such as \texttt{matsnu} (with a recall of only 0.158). \texttt{matsnu} is a dictionary-based DGA relying on a built-in list of more than 1 300 verbs and nouns. As the dataset used for the evaluation only contains 12 714 examples of \texttt{matsnu} domains, this was probably insufficient to learn the underlying regularities produced by this DGA.

%maybe say world-list based DGA relying on a hardcoded list of more than..

\section{Conclusion}
\label{sec:conclusion}

This paper presented a data-driven approach to the automatic detection of malware-generated domain names using recurrent neural networks. Although the idea of using recurrent neural networks for detecting malicious domains is not entirely new \citep{woodbridge2016predicting}, the paper is to our knowledge the first one to evaluate it on a large dataset of several million malware-generated domains (covering a total of 61 malware families). The model does not require any handcrafted feature and can be easily retrained to respond to new malwares. Furthermore, it can be directly applied on the raw domain names, without requiring access to additional contextual information. 

Future work will investigate the integration of this neural model as part of a larger machine-learning architecture for the detection of cyber-threats in traffic data.

\bibliography{biblio}

\begin{thebibliography}{22}
\providecommand{\natexlab}[1]{#1}
\providecommand{\url}[1]{\texttt{#1}}
\expandafter\ifx\csname urlstyle\endcsname\relax
  \providecommand{\doi}[1]{doi: #1}\else
  \providecommand{\doi}{doi: \begingroup \urlstyle{rm}\Url}\fi

\bibitem[Anderson et~al.(2016)Anderson, Woodbridge, and
  Filar]{anderson2016deepdga}
H.~S. Anderson, J.~Woodbridge, and B.~Filar.
\newblock {DeepDGA}: Adversarially-tuned domain generation and detection.
\newblock In \emph{Proceedings of the 2016 ACM Workshop on Artificial
  Intelligence and Security}, pages 13--21. ACM, 2016.

\bibitem[Antonakakis et~al.(2012)Antonakakis, Perdisci, Nadji, Vasiloglou,
  Abu-Nimeh, Lee, and Dagon]{antonakakis2012}
M.~Antonakakis, R.~Perdisci, Y.~Nadji, N.~Vasiloglou, S.~Abu-Nimeh, W.~Lee, and
  D.~Dagon.
\newblock From throw-away traffic to bots: Detecting the rise of {DGA}-based
  malware.
\newblock In \emph{USENIX security symposium}, volume~12, 2012.

\bibitem[Bahdanau et~al.(2014)Bahdanau, Cho, and Bengio]{bahdanau+al-2014-nmt}
D.~Bahdanau, K.~Cho, and Y.~Bengio.
\newblock Neural machine translation by jointly learning to align and
  translate.
\newblock \emph{arXiv e-prints}, abs/1409.0473, September 2014.

\bibitem[Barabosch et~al.(2012)Barabosch, Wichmann, Leder, and
  Gerhards-Padilla]{barabosch2012}
T.~Barabosch, A.~Wichmann, F.~Leder, and E.~Gerhards-Padilla.
\newblock Automatic extraction of domain name generation algorithms from
  current malware.
\newblock In \emph{Proc. NATO Symposium IST-111 on Information Assurance and
  Cyber Defense, Koblenz, Germany}, 2012.

\bibitem[Chollet et~al.(2015)]{chollet2015keras}
F.~Chollet et~al.
\newblock Keras.
\newblock \url{https://github.com/fchollet/keras}, 2015.

\bibitem[Chung et~al.(2014)Chung, G{\"{u}}l{\c{c}}ehre, Cho, and
  Bengio]{DBLP:journals/corr/ChungGCB14}
J.~Chung, {\c{C}}~G{\"{u}}l{\c{c}}ehre, K.~Cho, and Y.~Bengio.
\newblock Empirical evaluation of gated recurrent neural networks on sequence
  modeling.
\newblock \emph{CoRR}, abs/1412.3555, 2014.

\bibitem[Goldberg(2016)]{goldberg2016primer}
Y.~Goldberg.
\newblock A primer on neural network models for natural language processing.
\newblock \emph{Journal of Artificial Intelligence Research (JAIR)},
  57:\penalty0 345--420, 2016.

\bibitem[Goodfellow et~al.(2016)Goodfellow, Bengio, and
  Courville]{Goodfellow-et-al-2016}
I.~Goodfellow, Y.~Bengio, and A.~Courville.
\newblock \emph{Deep Learning}.
\newblock MIT Press, 2016.

\bibitem[Graves et~al.(2013)Graves, Mohamed, and
  Hinton]{DBLP:conf/icassp/GravesMH13}
A.~Graves, A.~Mohamed, and G.~E. Hinton.
\newblock Speech recognition with deep recurrent neural networks.
\newblock In \emph{{IEEE} International Conference on Acoustics, Speech and
  Signal Processing ({ICASSP} 2013)}, pages 6645--6649, 2013.

\bibitem[Grill et~al.(2015)Grill, Nikolaev, Valeros, and Rehak]{grill2015}
M.~Grill, I.~Nikolaev, V.~Valeros, and M.~Rehak.
\newblock Detecting {DGA} malware using netflow.
\newblock In \emph{Integrated Network Management (IM), 2015 IFIP/IEEE
  International Symposium on}, pages 1304--1309. IEEE, 2015.

\bibitem[Hochreiter and Schmidhuber(1997)]{Hochreiter:1997:LSM:1246443.1246450}
S.~Hochreiter and J.~Schmidhuber.
\newblock Long short-term memory.
\newblock \emph{Neural Computation}, 9\penalty0 (8):\penalty0 1735--1780,
  November 1997.
\newblock ISSN 0899-7667.

\bibitem[K{\"u}hrer et~al.(2014)K{\"u}hrer, Rossow, and Holz]{kuhrer2014paint}
M.~K{\"u}hrer, C.~Rossow, and T.~Holz.
\newblock Paint it black: Evaluating the effectiveness of malware blacklists.
\newblock In \emph{International Workshop on Recent Advances in Intrusion
  Detection}, pages 1--21. Springer, 2014.

\bibitem[Plohmann et~al.(2016)Plohmann, Yakdan, Klatt, Bader, and
  Gerhards-Padilla]{197187}
D.~Plohmann, K.~Yakdan, M.~Klatt, J.~Bader, and E.~Gerhards-Padilla.
\newblock A comprehensive measurement study of domain generating malware.
\newblock In \emph{25th {USENIX} Security Symposium ({USENIX} Security 16)},
  pages 263--278, 2016.

\bibitem[Ruder(2017)]{DBLP:journals/corr/Ruder17a}
S.~Ruder.
\newblock An overview of multi-task learning in deep neural networks.
\newblock \emph{CoRR}, abs/1706.05098, 2017.

\bibitem[Schiavoni et~al.(2014)Schiavoni, Maggi, Cavallaro, and
  Zanero]{schiavoni2014phoenix}
S.~Schiavoni, F.~Maggi, L.~Cavallaro, and S.~Zanero.
\newblock Phoenix: {DGA}-based botnet tracking and intelligence.
\newblock In \emph{International Conference on Detection of Intrusions and
  Malware, and Vulnerability Assessment}, pages 192--211, 2014.

\bibitem[Schuster and Paliwal(1997)]{Schuster:1997:BRN:2198065.2205129}
M.~Schuster and K.K. Paliwal.
\newblock Bidirectional recurrent neural networks.
\newblock \emph{IEEE Transactions on Signal Processing}, 45\penalty0
  (11):\penalty0 2673--2681, November 1997.

\bibitem[Villamarin-Salomon and Brustoloni(2008)]{villamarin2008identifying}
R.~Villamarin-Salomon and J.~C. Brustoloni.
\newblock Identifying botnets using anomaly detection techniques applied to
  {DNS} traffic.
\newblock In \emph{Consumer Communications and Networking Conference, 2008.
  CCNC 2008. 5th IEEE}, pages 476--481. IEEE, 2008.

\bibitem[Vinyals and Le(2015)]{vinyals2015neural}
O.~Vinyals and Q.~Le.
\newblock A {Neural} {Conversational} {Model}.
\newblock \emph{CoRR}, abs/1506.05869, 2015.

\bibitem[Woodbridge et~al.(2016)Woodbridge, Anderson, Ahuja, and
  Grant]{woodbridge2016predicting}
J.~Woodbridge, H.~S Anderson, A.~Ahuja, and D.~Grant.
\newblock Predicting domain generation algorithms with long short-term memory
  networks.
\newblock \emph{arXiv preprint arXiv:1611.00791}, 2016.

\bibitem[Yadav and Reddy(2011)]{yadav2011winning}
S.~Yadav and AL~N. Reddy.
\newblock Winning with {DNS} failures: Strategies for faster botnet detection.
\newblock In \emph{International Conference on Security and Privacy in
  Communication Systems}, pages 446--459. Springer, 2011.

\bibitem[Yadav et~al.(2010)Yadav, Reddy, Reddy, and Ranjan]{yadav2010detecting}
S.~Yadav, A.~K.~K. Reddy, AL~Reddy, and S.~Ranjan.
\newblock Detecting algorithmically generated malicious domain names.
\newblock In \emph{Proceedings of the 10th ACM SIGCOMM conference on Internet
  measurement}, pages 48--61. ACM, 2010.

\bibitem[Zhou et~al.(2013)Zhou, Li, Miao, and Yim]{zhou2013dga}
Y.~Zhou, Q.-S. Li, Q.~Miao, and K.~Yim.
\newblock {DGA}-based botnet detection using {DNS} traffic.
\newblock \emph{J. Internet Serv. Inf. Secur.}, 3\penalty0 (3/4):\penalty0
  116--123, 2013.

\end{thebibliography}

\end{document}